\documentclass[twocolumn]{aastex63} 

\usepackage{graphicx} 
\usepackage{float} 
\usepackage{wrapfig} 
\usepackage{lipsum} 
\usepackage{hyperref}
\usepackage{times}
\usepackage{amsmath}
\usepackage{graphicx}
\usepackage{subfigure}
\usepackage{placeins}
\usepackage{hyperref}
\usepackage{gensymb}
\usepackage{upgreek}
\usepackage{natbib}

\usepackage{graphicx}
\usepackage{subfigure}
\usepackage{multirow}
\usepackage{comment}
\usepackage{natbib}
\usepackage{hyperref}
\usepackage{mathtools}
\usepackage{mathrsfs}
\usepackage{fontenc}
\usepackage{color}
\usepackage{url}
\usepackage{hyperref}
\usepackage{gensymb}
\usepackage{pifont}
\graphicspath{{./}}

\newcommand       \K            {\,{\rm K}}
\newcommand       \mum          {{\rm \mu m}}

\submitjournal{ApJ}

\shorttitle{Contribution of Evolved Stars to PAH Heating}
\shortauthors{Zhang \& Ho}

\begin{document}

\title{The Contribution of Evolved Stars to PAH Heating and Implications for Estimating Star Formation Rates}

\author[0000-0003-4937-9077]{Lulu Zhang}
\affiliation{Kavli Institute for Astronomy and Astrophysics, Peking University, Beijing 100871, China; l.l.zhang@pku.edu.cn}
\affiliation{Department of Astronomy, School of Physics, Peking University, Beijing 100871, China}

\author[0000-0001-6947-5846]{Luis C. Ho}
\affiliation{Kavli Institute for Astronomy and Astrophysics, Peking University, Beijing 100871, China; l.l.zhang@pku.edu.cn}
\affiliation{Department of Astronomy, School of Physics, Peking University, Beijing 100871, China}


\begin{abstract}
Emission from polycyclic aromatic hydrocarbons (PAHs) is a promising tool for estimating star formation rate (SFR) in galaxies, but the origin of its sources of excitation, which include not only young but possibly also old stars, remains uncertain. We analyze Spitzer mid-infrared mapping-mode spectroscopic observations of the nuclear and extra-nuclear regions of 33 nearby galaxies to study the contribution of evolved stars to PAH emission. In combination with photometric measurements derived from ultraviolet, H$\alpha$, and infrared images, the spatially resolved spectral decomposition enables us to characterize the PAH emission, SFR, and stellar mass of the sample galaxies on sub-kpc scales. We demonstrate that the traditional empirical correlation between PAH luminosity and SFR has a secondary dependence on specific SFR, or, equivalently, stellar mass. Ultraviolet-faint regions with lower specific SFRs and hence greater fraction of evolved stars emit stronger PAH emission at fixed SFR than ultraviolet-bright regions. We reformulate the PAH-based SFR estimator by explicitly introducing stellar mass as a second parameter to account for the contribution of evolved stars to PAH excitation. The influence of evolved stars can explain the sub-linear correlation between PAH emission and SFR, and it can partly account for the PAH deficit in dwarf galaxies and low-metallicity environments.
\end{abstract}


\section{Introduction} 

Estimating the strength of star formation is of great importance for characterizing the formation and evolution of galaxies. Normal star-forming or starburst galaxies exhibit significant mid-infrared (IR) broad emission features (e.g., \citealt{Smith et al. 2007b, Galliano et al. 2008b, Gordon et al. 2008}), the carriers of which are widely attributed to polycyclic aromatic hydrocarbons (PAHs; \citealt{Duley & Williams 1981, Leger & Puget 1984, Allamandola et al. 1985}; see review in \citealt{Tielens 2008}). PAHs radiate through IR fluorescence following electronic excitation after absorbing a single ultraviolet (UV) photon \citep{Allamandola et al. 1989, Tielens 2005}. Thus, PAH emission provides an indirect estimate of the strength of the UV radiation field, and hence the star formation rate (SFR), which is relatively free from dust extinction \citep{Forster Schreiber et al. 2004, Peeters et al. 2004}. Accordingly, numerous studies have provided SFR prescriptions based on PAH emission (e.g., \citealt{Calzetti et al. 2005, Calzetti et al. 2007, Wu et al. 2005, Treyer et al. 2010, Shipley et al. 2016, Maragkoudakis et al. 2018, Xie & Ho 2019}). The utility of PAH emission as a SFR indicator has motivated different strategies for measuring PAH strength. Some employ broadband photometric observations that capture strong PAH features (e.g., \citealt{Wu et al. 2005, Calzetti et al. 2007, Donoso et al. 2012, Lee et al. 2013, Cluver et al. 2017, Mahajan et al. 2019, Zhang & Ho 2022}), which, while expedient, may be susceptible to uncertainties from the underlying continuum contamination (e.g., \citealt{Donoso et al. 2012, Crocker et al. 2013, Cluver et al. 2014, Lu et al. 2014}), while others have advanced methods to extract individual PAH features from spectroscopic observations, after subtracting the local continuum or simultaneously fitting the PAH features and the continuum using model decomposition (e.g., \citealt{Peeters et al. 2002, Marshall et al. 2007, Smith et al. 2007b, Xie et al. 2018, Zhang et al. 2021}). 

Although PAH emission is widely used as a tracer of star formation activity, the theoretical assumption that PAHs are primarily excited by ionizing UV photons from newborn stars, with little or no absorption of visible photons from old stars \citep{Sellgren 2001}, is not always valid \citep{Draine & Li 2001}. \cite{Li & Draine 2002} gave a satisfactory model fit to the observed IR emission of vdB\,133, a very UV-poor reflection nebula, demonstrating that astronomical PAHs can be pumped even by cool stars with weak UV radiation. \cite{Haas et al. 2002} studied the spatial coincidence between the strength of the PAH~7.7\,$\mum$ feature and the 850\,$\mum$ continuum emission for five galaxies with a range of star formation activity, and they suggested that PAHs are preferentially distributed in regions dominated by cold dust and molecular clouds, where PAHs are excited mainly by the interstellar radiation field (see also \citealt{Bendo et al. 2008, Bendo et al. 2010, Cortzen et al. 2019}). The excitation of PAHs may vary depending on galaxy type. For example, \cite{Jones et al. 2015} studied the correlation between the 8\,$\mum$ PAH emission band, as captured in the IRAC4 band of the Spitzer Space Telescope \citep{Houck et al. 2004}, and the dust continuum in NGC\,2403 and M\,83. They concluded that PAHs in NGC\,2403 are coupled with the diffuse dust and are heated by the radiation from the total stellar population, while the excitation of PAHs within M\,83 is connected more strongly with star-forming regions. Similarly, \cite{Bendo et al. 2020} studied the PAH excitation for a sample of 25 spiral galaxies by analyzing the ratio of PAH surface brightness to dust surface density and its correlation with tracers of young and evolved stars. They found that PAH emission is correlated straightforwardly with young stars (as traced by the UV+24\,$\mum$ continuum) in 11/25 galaxies, whereas for 5/25 galaxies PAH emission is linked with young stars only in the inner parts of the galaxies. Moreover, PAH emission better traces evolved stars (3.6\,$\mum$ continuum) in 6/25 galaxies; no clear trends can be discerned in the remaining 3/25 galaxies. Some case studies have tried to quantify the specific contribution of evolved stars to PAH excitation. For example, \cite{Crocker et al. 2013} concluded that $30\%-43\%$ of the 8\,$\mum$ band PAH emission of NGC\,628 is related to evolved stars, while \cite{Lu et al. 2014} proposed that 67\% of the 8\,$\mum$ band PAH emission in M\,81 is heated by evolved stars.

The aforementioned results testify that a single source cannot completely account for the excitation of PAHs in different galaxies, and even of different environments within a single galaxy. The contribution of evolved stars should be considered, to ascertain the degree to which it might influence the SFRs based on PAH emission. To this end, spatially resolved analysis is critical to help isolate the contribution of PAHs from different environments, in an attempt to identify potentially different sources of excitation. We perform this exercise using spatially resolved measurements of PAH emission extracted from galaxies from the Spitzer Infrared Nearby Galaxies Survey\footnote{\url{https://irsa.ipac.caltech.edu/data/SPITZER/SINGS/}} (SINGS; \citealt{Kennicutt et al. 2003}). We quantify the influence of evolved stars on PAH emission and propose a revised calibration of the PAH-based SFR estimator, which has promising applications for observations from the James Webb Space Telescope \citep{Rigby et al. 2022}.

\begin{figure}[t]
\center{\includegraphics[width=1\linewidth]{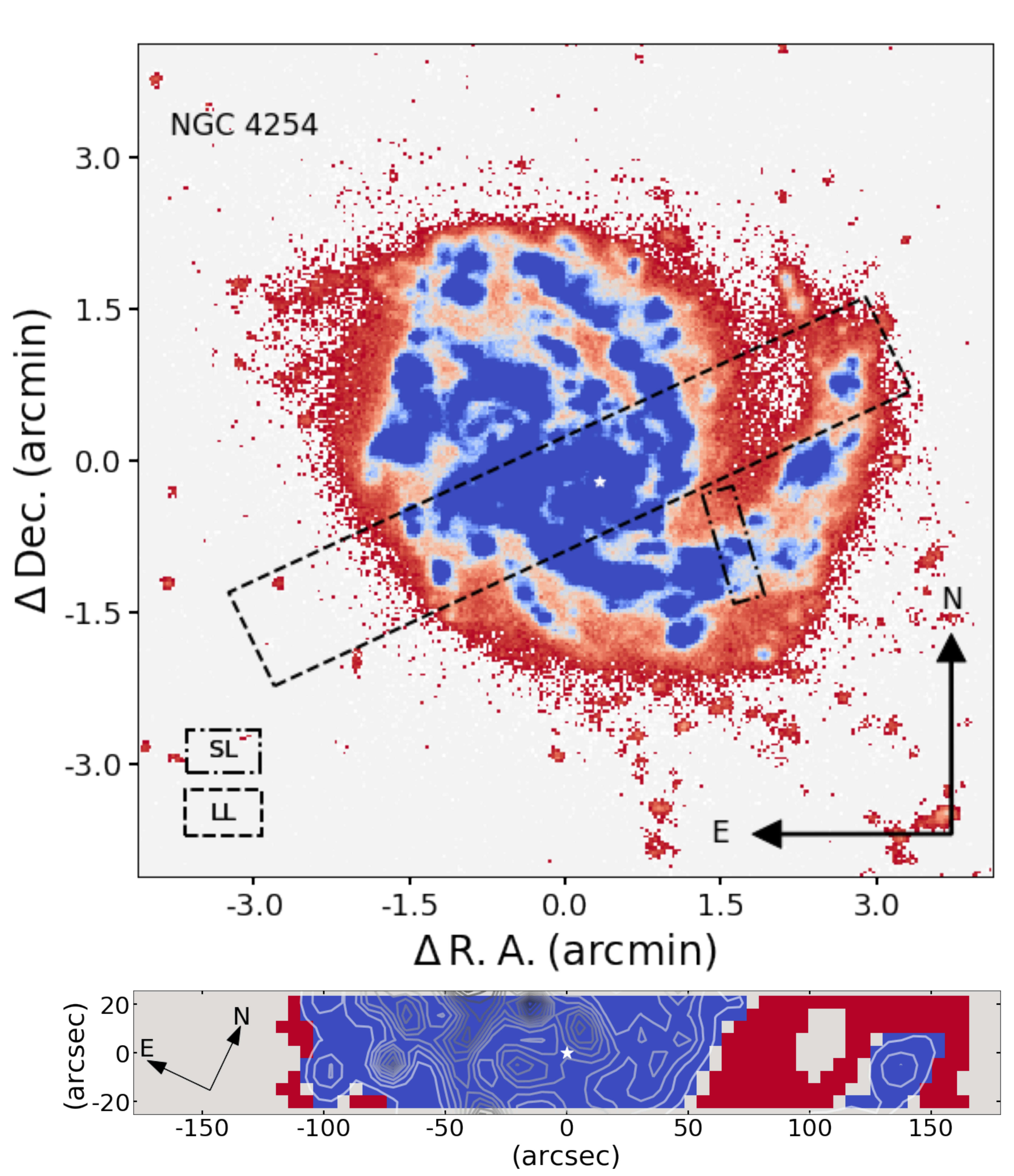}}
\caption{{Illustration of the spatial coverage of the IRS mapping-mode data cube using the FUV image of NGC\,4254 (top), whose center is marked with a star. The bottom panel delineates the UV-bright spaxels (blue) and UV-faint spaxels (red) within the LL data cube that bisects the galaxy; all the remaining spaxels are in gray. Analogous UV-bright and UV-faint spaxels can be defined for the SL data cube that samples the extranuclear H\,{\small II} region on the spiral arm.}\label{fig:coverage}}
\end{figure}

\begin{deluxetable}{cccccc}[!ht]
\tabletypesize{\small}
\tablecolumns{6}
\tablecaption{Multiwavelength Photometric Images}
\tablehead{
\colhead{Telescope} & \colhead{Filter} & \colhead{$\lambda_{\rm eff}$} & \colhead{FWHM} & \colhead{$\sigma_{\rm cal}$} & \colhead{References}\\
\colhead{} & \colhead{} & \colhead{$(\mu m)$} & \colhead{(\arcsec) } & \colhead{(\%)} & \colhead{} \\
\colhead{(1)} & \colhead{(2)} & \colhead{(3)} & \colhead{(4)} & \colhead{(5)} & \colhead{(6)} 
}
\startdata
GALEX & FUV & 0.153 & 4.2 & 5 & 1, 2\\
GALEX & NUV & 0.231 & 5.3 & 3 & 1, 2\\
KPNO/CTIO & H$\alpha$ & 0.657 & 1.9 & 15 & 3, 4\\
2MASS & $J$  & 1.25 & 2.5 & 5 & 5, 6\\
2MASS & $H$ & 1.65 & 2.5 & 5 & 5, 6\\
2MASS & $K_s$ & 2.16 & 2.5 & 5 & 5, 6\\
WISE &  W1 & 3.35 & 6.1 & 2.4 & 7, 8\\
WISE &  W2 & 4.60 & 6.4 & 2.8 & 7, 8\\
WISE &  W3 & 11.56 & 6.5 & 4.5 & 7, 8\\
Spitzer & IRAC1 & 3.55 & 1.7 & 10 & 3, 9\\
Spitzer & IRAC2 & 4.49 & 1.7 & 10 & 3, 9\\
Spitzer & IRAC3 & 5.73 & 1.9 & 10 & 3, 9\\
Spitzer & IRAC4 & 7.87 & 2.0 & 10 & 3, 9\\
Spitzer & MIPS24 & 23.68 & 6.0 & 4 & 3, 10\\
\enddata
\tablecomments{Col. (1): Telescope. Col. (2): Filter. Col. (3): Effective wavelength of filter. Col. (4): FWHM of the PSF. Col. (5): Calibration uncertainty.  Col. (6):  References.}
\tablerefs{(1) \citealt{Morrissey et al. 2007}; (2) \citealt{Gil de Paz et al. 2007}; (3) \citealt{Dale et al. 2007}; (4) \citealt{Kennicutt et al. 2008}; (5) \citealt{Jarrett et al. 2003}; (6) \citealt{Skrutskie et al. 2006}; (7) \citealt{Wright et al. 2010}; (8) \citealt{Jarrett et al. 2013}; (9) \citealt{Fazio et al. 2004}; (10) \citealt{Engelbracht et al. 2007}.}
\label{tab:tableimg}
\end{deluxetable}

\setlength{\tabcolsep}{3pt}
\begin{deluxetable*}{lcccccc||ccccccc}
\tabletypesize{\small}
\tablecolumns{14}
\tablecaption{Properties of the Galaxy Sample}
\tablehead{
\colhead{Galaxy} & \colhead{Morphology} & \colhead{Class} & \colhead{$D_L$} & \colhead{[N {\scriptsize II}]/H$\alpha$} & \colhead{$E(B-V)$} & \colhead{$N_{\rm H~II}$} & \colhead{Galaxy} & \colhead{Morphology} & \colhead{Class} & \colhead{$D_L$} & \colhead{[N {\scriptsize II}]/H$\alpha$} & \colhead{$E(B-V)$} & \colhead{$N_{\rm H~II}$}\\
\colhead{(1)} & \colhead{(2)} & \colhead{(3)} & \colhead{(4)} & \colhead{(5)} & \colhead{(6)} & \colhead{(7)} & \colhead{(1)} & \colhead{(2)} & \colhead{(3)} & \colhead{(4)} & \colhead{(5)} & \colhead{(6)} & \colhead{(7)}
}
\startdata
NGC\,24 & Sc & SFG & 8.2 & 0.37 & 0.017 & 0 & NGC\,3521 & SABbc & AGN & 11.2 & 0.56 & 0.050 & 4 \\
NGC\,337 & SBd & SFG & 19.3 & 0.23 & 0.096 & 0 & NGC\,3627 & SABb & AGN & 9.38 & 0.55 & 0.029 & 3 \\
NGC\,628 & Sc & SFG & 7.2 & 0.34 & 0.061 & 4 & NGC\,3938 & Sc & SFG & 17.9 & 0.42 & 0.018 & 1 \\
NGC\,855 & E & SFG & 9.73 & 0.18 & 0.061 & 0 & NGC\,4254 & Sc & SFG & 14.4 & 0.45 & 0.033 & 2 \\
NGC\,925 & SABd & SFG & 9.12 & 0.20 & 0.066 & 0 & NGC\,4321 & SABbc & AGN & 14.3 & 0.43 & 0.023 & 3 \\
NGC\,1097 & SBb & AGN & 14.2 & 0.69 & 0.023 & 3 & NGC\,4536 & SABbc & SFG & 14.5 & 0.45 & 0.016 & 0 \\
NGC\,1482 & S0 & SFG & 22.6 & 0.69 & 0.034 & 0 & NGC\,4559 & SABcd & SFG & 6.98 & 0.28 & 0.015 & 0 \\
NGC\,1566 & SABbc & AGN & 18.0 & 0.62 & 0.008 & 3 & NGC\,4625 & SABm & SFG & 9.3 & 0.53 & 0.016 & 0 \\
NGC\,2403 & SABcd & SFG & 3.5 & 0.22 & 0.034 & 6 & NGC\,4631 & SBd & SFG & 7.62 & 0.28 & 0.015 & 3 \\
NGC\,2798 & SBa & SFG & 25.8 & 0.36 & 0.017 & 0 & NGC\,4736 & Sab & AGN & 4.66 & 0.71 & 0.015 & 3 \\
NGC\,2915 & I0 & SFG & 3.78 & 0.14 & 0.236 & 0 & NGC\,5055 & Sbc & AGN & 7.94 & 0.49 & 0.015 & 1 \\
NGC\,2976 & Sc & SFG & 3.55 & 0.36 & 0.064 & 2 & NGC\,5194 & SABbc & AGN & 8.2 & 0.59 & 0.031 & 11 \\
NGC\,3034 & I0 & SFG & 3.50 & 0.30 & 0.134 & 0 & NGC\,5713 & SABbc & SFG & 21.4 & 0.55 & 0.034 & 2 \\
NGC\,3049 & SBab & SFG & 19.2 & 0.40 & 0.033 & 0 & NGC\,6946 & SABcd & SFG & 6.8 & 0.45 & 0.294 & 0 \\
NGC\,3184 & SABcd & SFG & 11.7 & 0.52 & 0.014 & 0 & NGC\,7552 & Sc & SFG & 22.3 & 0.43 & 0.012 & 0 \\
NGC\,3198 & SBc & SFG & 14.1 & 0.30 & 0.011 & 0 & NGC\,7793 & Sd & SFG & 3.91 & 0.31 & 0.017 & 4 \\
NGC\,3351 & SBb & SFG & 9.33 & 0.66 & 0.024 & 0 &   &   &   &   &   &   &   \\
\enddata
\tablecomments{Col. (1): Galaxy name. Col. (2): Optical morphology from the NASA/IPAC Extragalactic Database. Col. (3): Nuclear classification into SFG or AGN according to the [N\,{\footnotesize II}]/H$\alpha$ versus [O\,{\footnotesize III}]/H$\beta$ emission-line diagnostic diagram, based on spectra extracted from a central aperture of 2\farcs5$\times$2\farcs5 (\citealt{Moustakas et al. 2010}). Col. (4): Luminosity distance, in units of Mpc, from \cite{Dale et al. 2017}. Col. (5): [N\,{\footnotesize II}] $\lambda\lambda6548, 6584/$H$\alpha$ ratio adopted from \cite{Kennicutt et al. 2008}. Col. (6): $E(B-V)$, in units of mag, from \cite{Schlafly and Finkbeiner 2011}. Col. (7): Number of extra-nuclear H\,{\scriptsize II} regions in each galaxy.}
\label{tab:tablegal}
\end{deluxetable*}

\setlength{\tabcolsep}{2pt}
\begin{deluxetable*}{lrrrrrrrrrrrrrr}
\tabletypesize{\tiny}
\tablecolumns{15}
\tablecaption{Spatially Resolved Multiband Photometry}
\tablehead{
\colhead{Region} & \colhead{log $f_{\rm FUV}$} & \colhead{log $f_{\rm NUV}$} & \colhead{log $f_{\rm H\alpha}$} & \colhead{log $f_{J}$} & \colhead{log $f_{H}$} & \colhead{log $f_{Ks}$} & \colhead{log $f_{\rm W1}$} & \colhead{log $f_{\rm W2}$} & \colhead{log $f_{\rm W3}$} & \colhead{log $f_{\rm IRAC1}$} & \colhead{log $f_{\rm IRAC2}$} & \colhead{log $f_{\rm IRAC3}$} & \colhead{log $f_{\rm IRAC4}$} & \colhead{log $f_{\rm MIPS24}$}\\
\colhead{} & \colhead{(mJy)} & \colhead{(mJy)} & \colhead{(mJy)} & \colhead{(mJy)} & \colhead{(mJy)} & \colhead{(mJy)} & \colhead{(mJy)} & \colhead{(mJy)} & \colhead{(mJy)} & \colhead{(mJy)} & \colhead{(mJy)} & \colhead{(mJy)} & \colhead{(mJy)} & \colhead{(mJy)}\\
\colhead{(1)} & \colhead{(2)} & \colhead{(3)} & \colhead{(4)} & \colhead{(5)} & \colhead{(6)} & \colhead{(7)} & \colhead{(8)} & \colhead{(9)} & \colhead{(10)} & \colhead{(11)} & \colhead{(12)} & \colhead{(13)} & \colhead{(14)} & \colhead{(15)}
}
\startdata
NGC628\_UVB\_C & 0.09 $\pm$ 0.02 & 0.29 $\pm$ 0.01 & 1.37 $\pm$ 0.07 & 1.89 $\pm$ 0.03 & 1.95 $\pm$ 0.04 & 1.86 $\pm$ 0.02 & 1.58 $\pm$ 0.01 & 1.38 $\pm$ 0.01 & 1.90 $\pm$ 0.02 & 1.60 $\pm$ 0.04 & 1.43 $\pm$ 0.04 & 1.66 $\pm$ 0.04 & 2.02 $\pm$ 0.04 & 2.26 $\pm$ 0.02 \\
NGC628\_UVF\_C & 0.49 $\pm$ 0.02 & 0.72 $\pm$ 0.01 & 1.68 $\pm$ 0.07 & 2.43 $\pm$ 0.03 & 2.48 $\pm$ 0.03 & 2.39 $\pm$ 0.02 & 2.05 $\pm$ 0.01 & 1.85 $\pm$ 0.01 & 2.39 $\pm$ 0.02 & 2.10 $\pm$ 0.04 & 1.95 $\pm$ 0.04 & 2.09 $\pm$ 0.04 & 2.40 $\pm$ 0.04 & 2.42 $\pm$ 0.02 \\
NGC628\_UVB\_H1 & $-$0.87 $\pm$ 0.02 & $-$0.72 $\pm$ 0.01 & 0.76 $\pm$ 0.07 & 0.52 $\pm$ 0.05 & 0.48 $\pm$ 0.10 & 0.46 $\pm$ 0.03 & 0.33 $\pm$ 0.01 & 0.24 $\pm$ 0.01 & 1.09 $\pm$ 0.02 & 0.33 $\pm$ 0.04 & 0.36 $\pm$ 0.04 & 0.95 $\pm$ 0.04 & 1.19 $\pm$ 0.04 & 1.88 $\pm$ 0.02 \\
NGC628\_UVF\_H1 & $-$0.80 $\pm$ 0.02 & $-$0.55 $\pm$ 0.01 & 0.76 $\pm$ 0.07 & 1.03 $\pm$ 0.04 & 1.01 $\pm$ 0.06 & 0.90 $\pm$ 0.03 & 0.73 $\pm$ 0.01 & 0.55 $\pm$ 0.01 & 1.34 $\pm$ 0.02 & 0.78 $\pm$ 0.04 & 0.58 $\pm$ 0.04 & 0.92 $\pm$ 0.04 & 1.37 $\pm$ 0.04 & 1.57 $\pm$ 0.02 \\
NGC628\_UVB\_H2 & $-$0.81 $\pm$ 0.02 & $-$0.66 $\pm$ 0.01 & 0.71 $\pm$ 0.07 & 0.14 $\pm$ 0.13 & 0.03 $\pm$ 0.32 & $-$0.02 $\pm$ 0.10 & 0.12 $\pm$ 0.01 & 0.04 $\pm$ 0.01 & 0.96 $\pm$ 0.02 & 0.15 $\pm$ 0.05 & 0.21 $\pm$ 0.04 & 0.85 $\pm$ 0.04 & 1.12 $\pm$ 0.04 & 1.65 $\pm$ 0.02 \\
NGC628\_UVF\_H2 & $-$0.94 $\pm$ 0.02 & $-$0.72 $\pm$ 0.01 & 0.66 $\pm$ 0.07 & 0.40 $\pm$ 0.12 & 0.24 $\pm$ 0.34 & \nodata & 0.17 $\pm$ 0.01 & $-$0.03 $\pm$ 0.01 & 1.07 $\pm$ 0.02 & 0.36 $\pm$ 0.05 & 0.15 $\pm$ 0.05 & 0.59 $\pm$ 0.04 & 1.14 $\pm$ 0.04 & 1.37 $\pm$ 0.02 \\
NGC628\_UVB\_H3 & $-$0.11 $\pm$ 0.02 & $-$0.01 $\pm$ 0.01 & 1.26 $\pm$ 0.07 & 0.52 $\pm$ 0.09 & 0.46 $\pm$ 0.19 & 0.16 $\pm$ 0.10 & 0.43 $\pm$ 0.01 & 0.33 $\pm$ 0.01 & 1.24 $\pm$ 0.02 & 0.47 $\pm$ 0.04 & 0.44 $\pm$ 0.04 & 0.97 $\pm$ 0.04 & 1.33 $\pm$ 0.04 & 1.80 $\pm$ 0.02 \\
NGC628\_UVF\_H3 & $-$0.93 $\pm$ 0.02 & $-$0.73 $\pm$ 0.01 & 0.58 $\pm$ 0.07 & 0.14 $\pm$ 0.18 & 0.08 $\pm$ 0.41 & \nodata & $-$0.02 $\pm$ 0.01 & $-$0.21 $\pm$ 0.01 & 0.82 $\pm$ 0.02 & 0.16 $\pm$ 0.05 & $-$0.04 $\pm$ 0.05 & 0.38 $\pm$ 0.04 & 0.92 $\pm$ 0.04 & 0.95 $\pm$ 0.02 \\
NGC628\_UVB\_H4 & $-$1.19 $\pm$ 0.02 & $-$1.10 $\pm$ 0.01 & 0.48 $\pm$ 0.07 & $-$0.67 $\pm$ 0.51 & $-$0.50 $\pm$ 0.68 & \nodata & $-$0.56 $\pm$ 0.01 & $-$0.62 $\pm$ 0.01 & 0.21 $\pm$ 0.02 & $-$0.53 $\pm$ 0.06 & $-$0.40 $\pm$ 0.05 & 0.23 $\pm$ 0.05 & 0.40 $\pm$ 0.05 & 0.90 $\pm$ 0.02 \\
NGC628\_UVF\_H4 & $-$1.09 $\pm$ 0.02 & $-$0.89 $\pm$ 0.01 & 0.66 $\pm$ 0.07 & $-$0.67 $\pm$ 1.17 & $-$0.38 $\pm$ 1.19 & \nodata & $-$0.39 $\pm$ 0.02 & $-$0.52 $\pm$ 0.02 & 0.64 $\pm$ 0.02 & $-$0.09 $\pm$ 0.06 & $-$0.24 $\pm$ 0.06 & 0.32 $\pm$ 0.05 & 0.73 $\pm$ 0.04 & 1.00 $\pm$ 0.02 \\
\enddata
\tablecomments{\footnotesize Spatially resolved multiband flux densities and corresponding uncertainties. (This table is available in its entirety in machine-readable form.)}
\label{tab:TablePhotI}
\end{deluxetable*}

\section{Observational Material}\label{section:sec2}

\subsection{Sample and Data Sets}

This work is based on the Spitzer and ancillary data from SINGS \citep{SINGS 2020}, a comprehensive IR imaging and spectroscopic survey that provides mapping-mode Infrared Spectrograph (IRS; \citealt{Werner et al. 2004}) observations of the central regions of 75 nearby galaxies, as well as of the extra-nuclear H\,{\small II} regions of a subset of 23 galaxies therein. The IRS has both low-resolution and high-resolution modules, each of which includes two modes, namely the short-low/long-low (SL/LL) modes and the short-high/long-high (SH/LH) modes. Each mode of the low-resolution module further consists of two orders (SL1/SL2 and LL1/LL2). Mapping-mode observations are conducted by scanning the slit of each mode to map a section of a galaxy. This paper only uses the low-resolution data, which cover the wavelength range $\sim 5-38\ \mum$ with a full width at half maximum (FWHM) spectral resolution of $\lambda/\Delta\lambda\approx 64-128$, depending on the observing mode. Our analysis is based on spectral data cubes constructed using {\tt CUBISM} \citep{Smith et al. 2007a}.

Figure~\ref{fig:coverage} schematically illustrates the spatial coverage of the observations. The LL mode data cubes, covering $14-38\ \mum$, map an area $50\arcsec \times 300\arcsec$ across the nuclear region of each galaxy with a point-spread function (PSF) of ${\rm FWHM} \approx 4\farcs5-9\farcs0$. The SL mode, capturing $5-14\ \mum$, samples a more limited area of $15\arcsec \times 50\arcsec$ with ${\rm FWHM} \approx 1\farcs5-3\farcs5$ resolution centered on extra-nuclear H\,{\small II} regions. Altogether, the spatially resolved spectra of the nuclear regions and H\,{\small II} regions provide a broad census of the behavior of PAH emission across a diverse range of environments. The large dynamic range in star-forming activity permits us to disentangle the contribution of evolved stars to the excitation of PAH emission.

None of the mapping-mode data cubes provides sufficient, continuous spectral coverage that can simultaneously measure the principal PAH features at 6.2, 7.7, 8.6, 11.3, and 17.0\,$\mum$. The central region of each galaxy was mapped only in the LL mode, while the extra-nuclear regions were confined to the SL mode. \cite{Zhang et al. 2021} devised a method that supplements the gaps in spectral coverage with mid-IR photometric measurements from WISE \citep{Wright et al. 2010, IPAC 2020} and Spitzer IRAC \citep{Fazio et al. 2004} and MIPS \citep{Rieke et al. 2004}. The combination of the photometry and partial spectroscopy yields a spectral decomposition that can produce a robust measurement of the integrated $5-20\ \mum$ PAH flux, as well as the strengths of the individual PAH features. \cite{Zhang et al. 2022} further extended the method to deliver estimates of the stellar mass by adding the near-IR bands of 2MASS \citep{Skrutskie et al. 2006}. Table~\ref{tab:tableimg} summarizes the multiwavelength imaging data sets used in this paper, which for the SINGS galaxies include images from the 2MASS Large Galaxy Atlas\footnote{\url{https://irsa.ipac.caltech.edu/data/LGA/}} \citep{Jarrett et al. 2003, Jarrett et al. 2020}, the WISE Atlas\footnote{\url{https://irsa.ipac.caltech.edu/applications/wise/}} \citep{Wright et al. 2010, IPAC 2020}, and MIPS24 \citep{Engelbracht et al. 2007}. We use H$\alpha$ imaging\footnote{\url{https://irsa.ipac.caltech.edu/data/SPITZER/SINGS/doc/sings_fifth_delivery_v2.pdf}} to set the benchmark for star formation activity, after correcting for contamination to the narrowband filter from [N\,{\small II}]\,$\lambda\lambda6548,6584$ using the [N\,{\small II}]/H$\alpha$ ratios of \cite{Kennicutt et al. 2008}. We additionally use broadband far-ultraviolet (FUV; 1350 \AA--1750 \AA) and near-ultraviolet (NUV; 1750 \AA--2750 \AA) images from the GALEX Ultraviolet Atlas of Nearby Galaxies\footnote{\url{https://archive.stsci.edu/prepds/galex_atlas/}} (\citealt{Gil de Paz et al. 2007}) to define two categories of regions (UV-bright and UV-faint), as illustrated in the bottom panel of Figure~\ref{fig:coverage} and described in Section~\ref{section:sec2.2}. 

The above data requirements can be fulfilled by the central regions of 25 SFGs and 55 extra-nuclear H\,{\small II} regions drawn from 10 star-forming galaxies (SFGs) and 8 galaxies hosting active galactic nuclei (AGNs) in the parent sample of SINGS, as listed in Table~\ref{tab:tablegal}. We have excluded four dwarf galaxies (IC\,4710, NGC\,1705, NGC\,6822, and Tol\,89), whose PAHs are likely systematically depleted by photo-destruction in the hard radiation field of low-metallicity environments (\citealt{Xie & Ho 2019, Zhang et al. 2022}).

\subsection{Data Processing and Definition of UV-bright and UV-faint Regions}\label{section:sec2.2}

The ancillary multi-wavelength images summarized in Table~1 form an integral part of our analysis, both in providing information on star formation (UV and H$\alpha$) and stellar mass ($JHK_s$), as well as supplementing the incomplete spectral coverage of the IRS spectra to enable spectral decomposition of the PAH emission. For the bands that display evident background residuals (i.e., H$\alpha$, 2MASS, WISE, IRAC and MIPS), we remove the artifacts by fitting and subtracting a two-dimensional polynomial function after masking out the real sources. The spatially resolved analysis of the IRS mapping-mode observations is performed using the method developed by \cite{Zhang et al. 2021, Zhang et al. 2022}, as summarized below.

To enable spatially resolved analysis, each IRS data cube and its respective ancillary images are convolved to the same angular resolution, which is set to ${\rm FWHM} = 6\farcs5$ for the SL data and ${\rm FWHM} = 10\farcs0$ for the LL data. These resolutions are chosen to be slightly larger than the largest PSF of the slices of the spectral data cubes (SL for the extra-nuclear H\,{\small II} regions; LL for the nuclear regions) and the relevant ancillary images (Table~\ref{tab:tableimg}). We perform the convolution using the {\tt convolve\_fft} function in {\tt astropy.convolution} (\citealt{Astropy Collaboration et al. 2013}), using convolution kernels constructed following \cite{Aniano et al. 2011}. All the convolved images are then reprojected into the same coordinate frame as the IRS data cubes, which has a pixel size of 1\farcs8 for the SL mode and 5\farcs1 for the LL mode. The spatially resolved multiband photometry is given in Table~\ref{tab:TablePhotI}. All the extracted IRS spectra are scaled to mitigate against the systematics produced by the convolution of the spectral data cube. The scaling factor for each integrated spectrum is determined individually by minimizing the weighted difference between the observed flux densities and those synthesized from the integrated IRS spectrum using the mid-IR bands that encompass the corresponding spectral range.

The main goal of this paper is to investigate whether evolved stars participate in the excitation of PAH emission, and, if so, to devise a means to quantify and remove its contribution from the PAH-based SFR estimator. Toward this end, we classify the spaxels into two categories according to their level of star formation activity, as judged by their UV brightness in the GALEX images: (1) UV-bright spaxels correspond to those that have flux densities that are $> 20\,\sigma$, where $\sigma$ is the background noise of the FUV and NUV images, and (2) UV-faint spaxels have flux densities between 3\,$\sigma$ and 20\,$\sigma$. The threshold between the two categories is, to some degree, arbitrary, but careful examination reveals that our choice effectively separates obvious, intense star-forming regions associated with spiral arms on the one hand, and more diffuse, less active inter-arm regions on the other hand. We use UV emission instead of H$\alpha$ emission as the benchmark because the UV continuum traces star formation activity on longer timescales. Moreover, in the case of H\,{\small II} regions in AGN host galaxies, we wish to avoid potential contamination from nebular emission from the AGN-photoionized narrow-line region, which can extend to sizable portions of the galactic disk (e.g., \citealt{Molina et al. 2022}). 

Then, we merge all the UV-bright spaxels of the nuclear region into a combined SED of high signal-to-noise ratio, and similarly for the UV-faint spaxels. The same procedure is applied to each of the extra-nuclear H\,{\small II} regions, if present. In the end, we have at least one UV-bright and one UV-faint SED for each galaxy (the exact number depends on the number of extra-nuclear H\,{\small II} regions, which varies from 0 to 11; see Table~2), which represent, respectively, the SEDs for regions with minimal versus substantial contributions from evolved stars.

\begin{figure}[t]
\center{\includegraphics[width=0.95\linewidth]{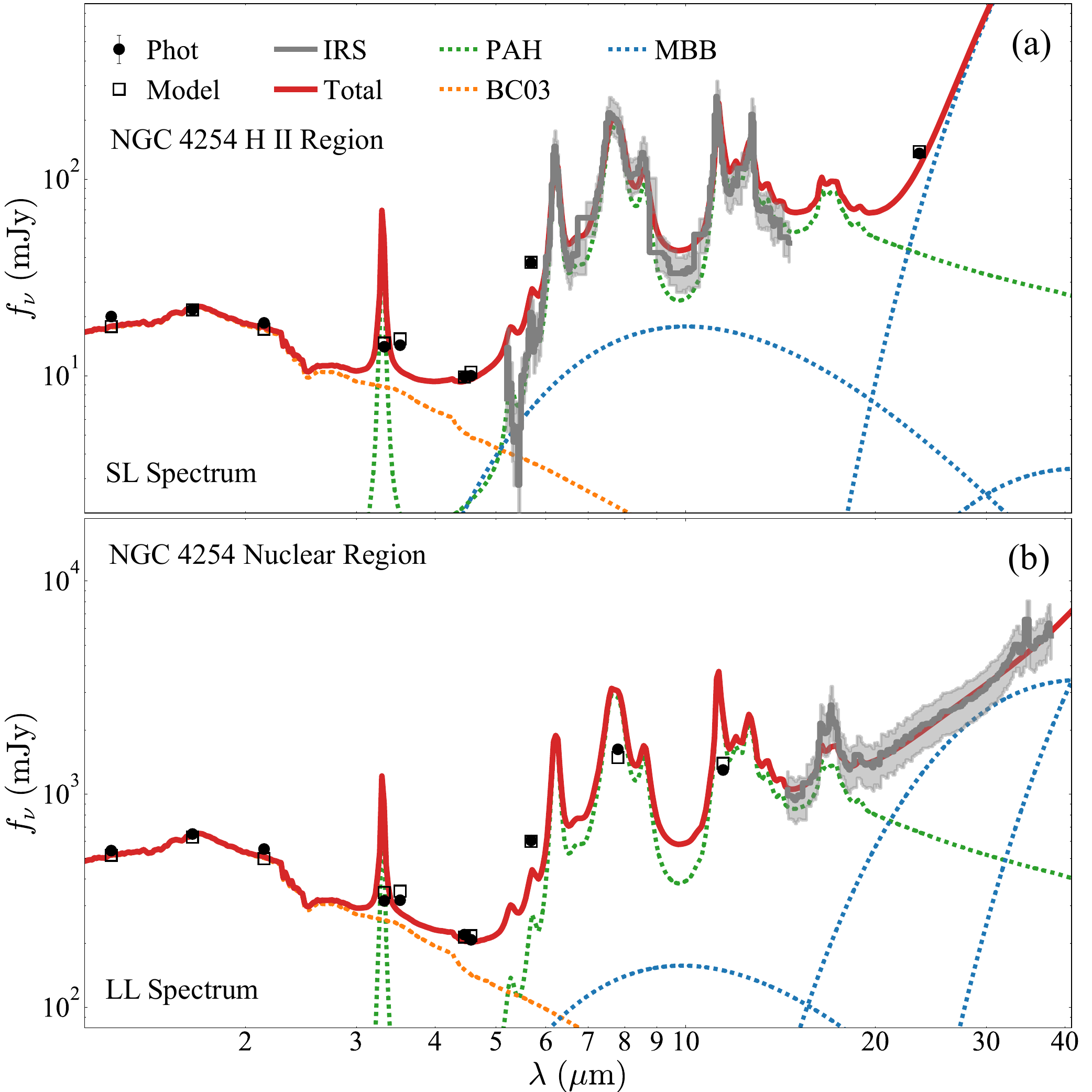}}
\caption{{Illustration of the PAH decomposition for the SFG NGC\,4254, showing (a) an SL spectrum of a spaxel from an H\,{\small II} region and (b) an LL spectrum of a spaxel from the nuclear region. The gray solid line is the IRS spectrum, and the black points are the photometric measurements. The red solid line is the best-fit model, which is composed of a PAH template, a stellar continuum component (\citealt{Bruzual & Charlot 2003}; BC03), and three dust continuum components each represented by a modified blackbody (MBB).}\label{fig:SED_Fitting}}
\end{figure}

\subsection{PAH, Stellar Mass, and SFR Measurements}\label{section:sec2.4}

Our method for measuring PAH is based on the technique of \cite{Xie et al. 2018}. After masking the strong ionic emission lines that are not blended with the main PAH features, we fit the $\sim 5-38\, \mum$ spectrum with a multicomponent model consisting of a theoretical PAH template plus dust continuum represented by three modified blackbodies of different temperatures, all subject to dust attenuation by foreground extinction. This method can decompose robustly the complex PAH features from low-resolution IRS spectra, as has been applied extensively in wide range of galactic environments, including the SINGS galaxies \citep{Xie & Ho 2019, Xie & Ho 2022, Zhang et al. 2022}. \cite{Zhang et al. 2021} modified the technique to enable reliable PAH measurement even when only partial low-resolution IRS spectra are available, by combining either the SL or LL spectra with properly matched mid-IR photometry to complement the missing spectral region. This improvement is essential to take advantage of the mapping-mode observations of SINGS, which lack full spectral coverage for most of the mapped regions of their sample galaxies. \cite{Zhang et al. 2022} further extended the SED coverage into the near-IR using the $JHK_s$ 2MASS photometric data in order to estimate the stellar mass. The stellar component is represented by a stellar population model from \cite{Bruzual & Charlot 2003}, assuming solar metallicity, a \cite{Chabrier 2003} stellar initial mass function, and a fixed stellar age of 6~Gyr, which is otherwise difficult to solve using IR information alone. Fortunately, the choice of stellar age does not affect our conclusions because the near-IR stellar emission is governed mostly by evolved stars, which have similar SEDs in the near-IR. 

Figure~\ref{fig:SED_Fitting} illustrates how the SED is decomposed by the multicomponent model. We use the Bayesian Markov chain Monte Carlo procedure {\tt emcee} in the {\tt Python} package to determine the posterior distribution of each best-fit parameter (see Section~4.2 in \citealt{Shangguan et al. 2018} for more details). The median and standard deviation of the posterior distribution of each best-fit parameter are taken as the final estimate and corresponding uncertainty, respectively. Following extensive experiments performed in \cite{Zhang et al. 2021}, we add an extra 20\% uncertainty to the IRS spectrum to balance the weighting of the spectroscopic and photometric data. The PAH luminosity, $L_{\rm PAH}$, listed in Table~\ref{tab:TableSFR} pertains to the integral of the best-fitting PAH template over the $5 - 20\ \mum$ region. The integrated PAH emission is less affected by the potential variation of individual PAH features due to variations in radiation intensity, which, in any case, are not strong for the range of conditions sampled by the nearby galaxies studied here \citep{Xie et al. 2018}. The strength of individual PAH features is strongly correlated with the strength of the integrated PAH emission (see, e.g., Table~4 in \citealt{Zhang et al. 2022}), but in this study we will not consider them further.

We use H$\alpha$ emission to quantify the strength of star formation activity, adopting the SFR indicator given by \cite{Kennicutt & Evans 2012}. To correct for foreground extinction, we use the Galactic extinction curve of \cite{Cardelli et al. 1989} with $R_V = 3.1$, for which $A_{\rm H\alpha} = 2.535\, E(B-V)$ \citep{Kennicutt et al. 2007}, and reddening values from \cite{Schlafly and Finkbeiner 2011}. Correction for internal extinction follows the prescription of \cite{Calzetti et al. 2007} for resolved H\,{\small II} regions, $L_{\rm H\alpha} = L_{\rm H\alpha}^{\rm obs} + 0.031\,L_{\rm 24\,\mum}$, where $L_{\rm H\alpha}^{\rm obs}$ is the Galactic extinction-corrected H$\alpha$ luminosity and $L_{\rm 24\,\mum}$ is derived from MIPS24 images. The spatially resolved measurements of PAH luminosity, ${\rm SFR} = 5.3\times\,10^{-42}L_{\rm H\alpha}$, and stellar mass are provided in Table~\ref{tab:TableSFR}.

\begin{deluxetable}{lccc}
\tabletypesize{\footnotesize}
\tablecolumns{4}
\tablecaption{Spatially Resolved Measurements}
\tablehead{
\colhead{Region} & \colhead{log $L_{\rm PAH}$} & \colhead{log $L_{\rm H\alpha}$}  & \colhead{log $M_{\rm \star}$} \\
\colhead{} & \colhead{($\rm erg\ s^{-1}$)} & \colhead{($\rm erg\ s^{-1}$)} & \colhead{($ M_{\odot}$)}\\
\colhead{(1)} & \colhead{(2)} & \colhead{(3)} & \colhead{(4)}
}
\startdata
NGC628\_UVB\_C & 41.22 $\pm$ 0.09 & 39.79 $\pm$ 0.11 & 8.75 $\pm$ 0.13 \\
NGC628\_UVF\_C & 41.62 $\pm$ 0.10 & 40.00 $\pm$ 0.10 & 9.27 $\pm$ 0.13 \\
NGC628\_UVB\_H1 & 40.39 $\pm$ 0.09 & 39.36 $\pm$ 0.13 & 7.29 $\pm$ 0.13 \\
NGC628\_UVF\_H1 & 40.57 $\pm$ 0.09 & 39.12 $\pm$ 0.11 & 7.83 $\pm$ 0.13 \\
NGC628\_UVB\_H2 & 40.36 $\pm$ 0.09 & 39.16 $\pm$ 0.11 & 6.82 $\pm$ 0.14 \\
NGC628\_UVF\_H2 & 40.36 $\pm$ 0.09 & 38.96 $\pm$ 0.10 & 6.32 $\pm$ 0.25 \\
NGC628\_UVB\_H3 & 40.59 $\pm$ 0.09 & 39.46 $\pm$ 0.08 & 7.20 $\pm$ 0.14 \\
NGC628\_UVF\_H3 & 40.14 $\pm$ 0.09 & 38.69 $\pm$ 0.07 & 6.10 $\pm$ 0.25 \\
NGC628\_UVB\_H4 & 39.71 $\pm$ 0.09 & 38.62 $\pm$ 0.07 & 6.08 $\pm$ 0.16 \\
NGC628\_UVF\_H4 & 40.00 $\pm$ 0.09 & 38.76 $\pm$ 0.07 & 5.67 $\pm$ 0.19 \\
\enddata
\tablecomments{\footnotesize Col. (1): Resolved region. Col. (2): Integrated ($5-20\ \mum$) PAH luminosity. Col. (3): Extinction-corrected H$\alpha$ luminosity. Col. (4): Stellar mass. (This table is available in its entirety in machine-readable form.) }
\label{tab:TableSFR}
\end{deluxetable}

\begin{figure*}[!ht]
\center{\includegraphics[width=0.975\textwidth]{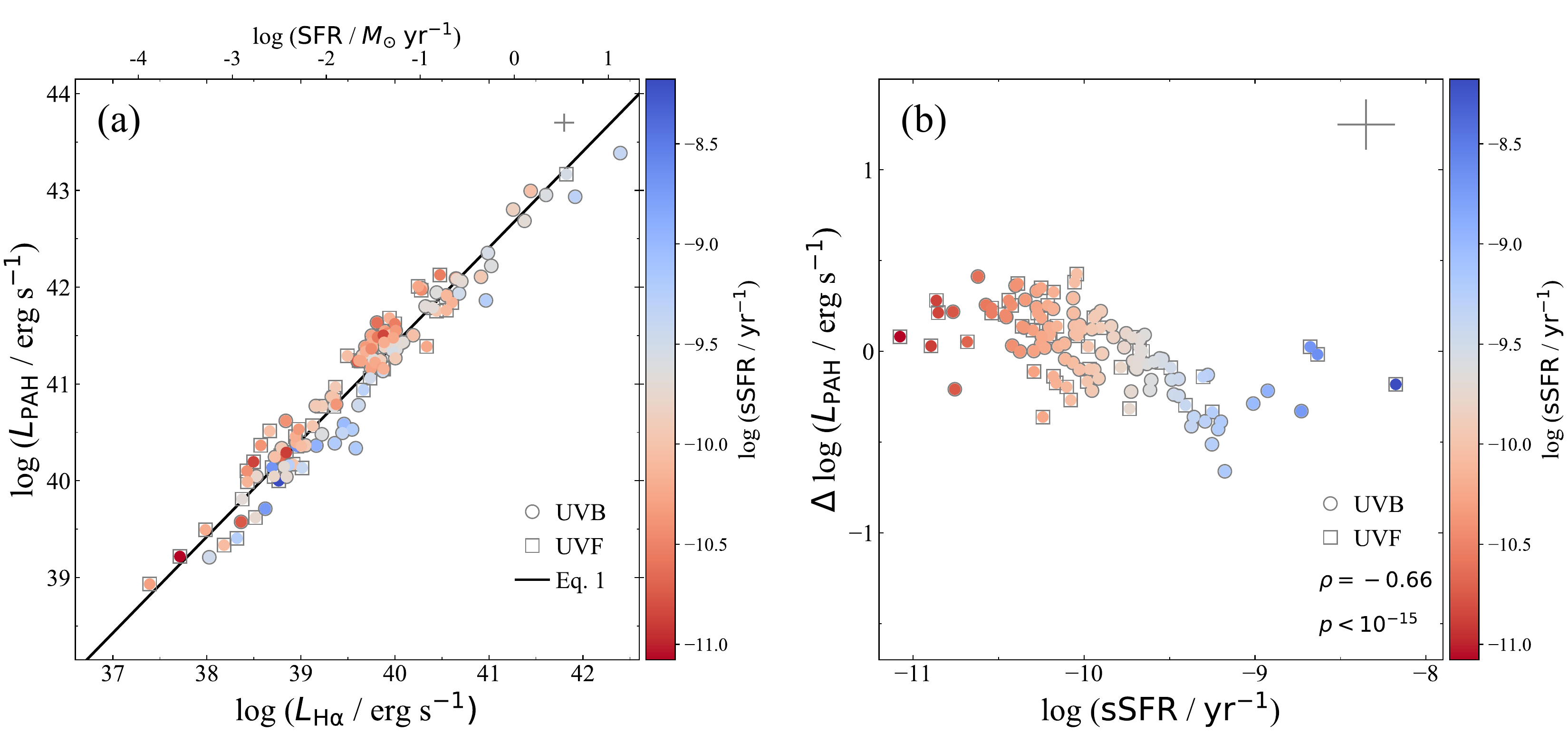}}
\caption{{(a) The correlation between integrated PAH luminosity and extinction-corrected H$\alpha$ luminosity for the UV-bright (UVB) and UV-faint (UVF) regions, color-coded according to the sSFR. The black solid line is the best linear regression fit for both regions combined (Equation~\ref{equ:SubRs}). (b) The dependence of the residual PAH luminosity between the observed PAH luminosity and the best-fit value (the black solid line in panel a) on the sSFR, color-coded according to the sSFR. The lower-right shows the Spearman correlation coefficient {$\rho$} and its statistical significance $p$. The median uncertainty is shown in the upper-right corner of each panel.}\label{fig:PAH_Ha_SF_I_II}}
\end{figure*}

\begin{figure}[t]
\center{\includegraphics[width=0.975\linewidth]{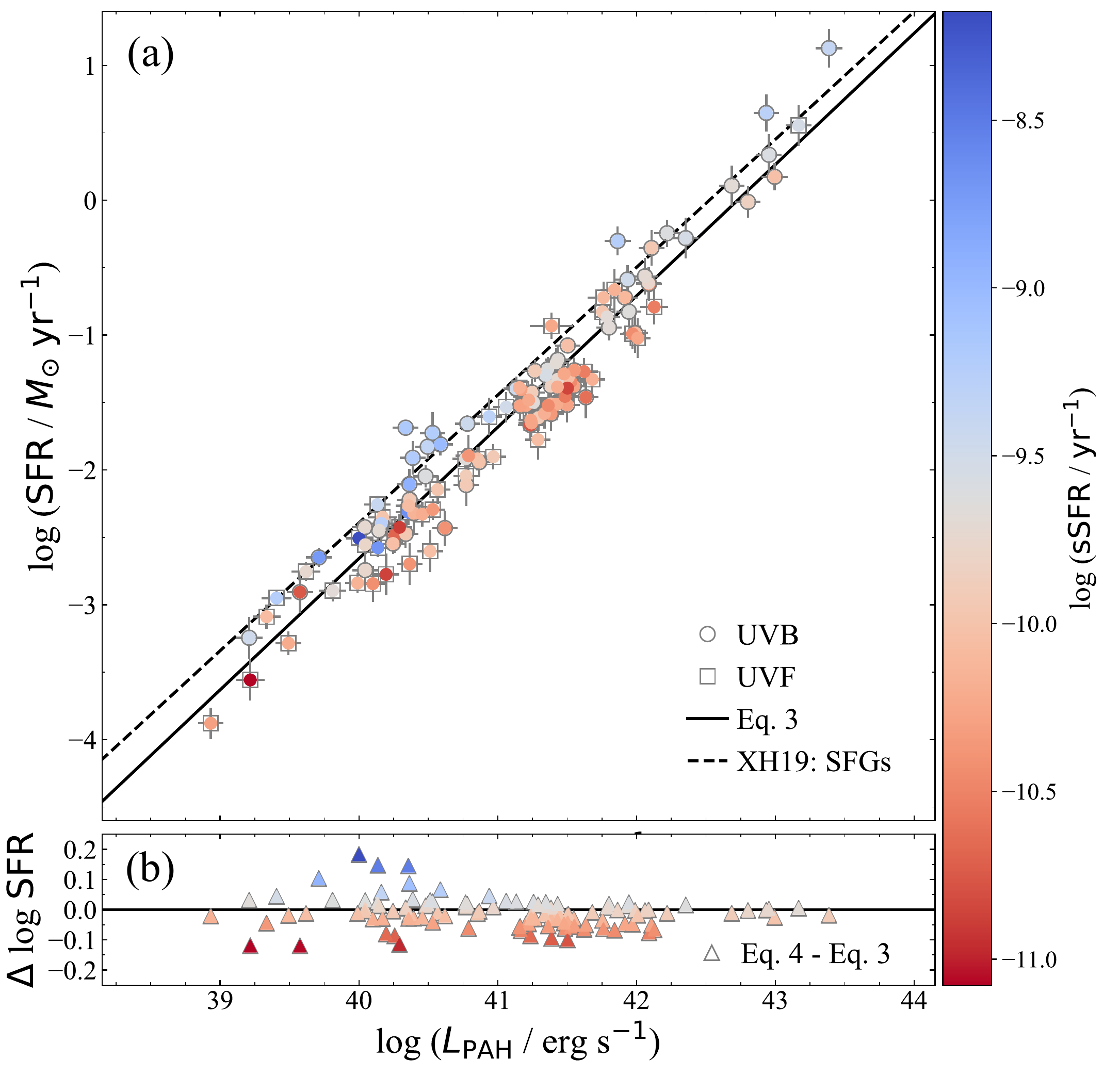}}
\caption{{(a) The correlation between observed SFR and PAH luminosity for the UV-bright (UVB) and UV-faint (UVF) regions, color-coded according to the sSFR. The black solid line is the direct calibration between observed SFR and PAH luminosity (Equation~\ref{equ:SFR_PAH}), and the black dashed line is the calibration for the sample of SFGs studied by \cite{Xie & Ho 2019}. (b) The difference between the revised SFR estimates (Equation~4) and the direct SFR estimates (Equation~3).}\label{fig:Rev_SFR}}
\end{figure}

\section{Results}\label{section:sec3}

\subsection{Contribution of Evolved Stars to PAH Emission}\label{sec:sec3.1}

We begin by examining the correlation between PAH luminosity and extinction-corrected H$\alpha$ luminosity, or, equivalently, SFR. For UV-bright and UV-faint regions combined, PAH emission tightly correlates with H$\alpha$ emission (Figure~\ref{fig:PAH_Ha_SF_I_II}a). A linear regression fit with the {\tt Python} package of the Bayesian method {\tt linmix} (\citealt{Kelly 2007}) gives

\noindent\small\begin{align}\label{equ:SubRs}
\begin{aligned}
&{\rm log}\,L_{\rm PAH} = (0.99\pm0.024)({\rm log}\, L_{\rm H\alpha} - 39.5) + (40.92\pm0.020),
\end{aligned}
\end{align}

\normalsize\noindent
with a total scatter of $\epsilon_t = 0.21$\,dex and an intrinsic scatter of $\epsilon_i = 0.16$\,dex. The total scatter is obtained as the root-mean-square error of the best-fit, and the intrinsic scatter is calculated assuming that the observed PAH luminosity is the dependent variable. Color coding the points by the specific SFR (${\rm sSFR \equiv SFR}/M_*$) reveals a subtle, but distinctive pattern. We observe that sSFR increases from the top-left to the lower-right of Figure~\ref{fig:PAH_Ha_SF_I_II}a, perpendicular to the correlated points. At a given strength of star formation activity, UV-faint regions, which have characteristically lower sSFR and hence a higher fraction of evolved stars, show stronger PAH emission than UV-bright regions, which tend to have higher values of sSFR. Figure~\ref{fig:PAH_Ha_SF_I_II}b highlights this trend more explicitly by plotting as a function of sSFR the residual difference between the observed PAH luminosity and the PAH luminosity predicted by Equation~\ref{equ:SubRs}. According to Figure~\ref{fig:PAH_Ha_SF_I_II}b, there is a strong inverse correlation between the relative PAH excess and sSFR: the Spearman correlation coefficient is $\rho = -0.66$, and the corresponding statistical significance parameter is $p < 10^{-15}$.

The above analysis shows that at given a SFR, star-forming regions with lower sSFR, namely those that contain more evolved stars, exhibit relatively stronger PAH emission. We suggest that this reflects the contribution of evolved stars to PAH heating. Although massive star formation activity can destroy PAHs and result in PAH suppression \citep{Gordon et al. 2008, Lebouteiller et al. 2011}, note that weaker PAH emission is found in regions with higher sSFR, rather than in those with higher SFR (Figure~\ref{fig:Rev_SFR}). On average, UV-faint regions ($\langle \log {\rm sSFR}\rangle = -10.03$) have a positive residual PAH luminosity ($\langle \Delta \log L_{\rm PAH}\rangle = +0.049$) while UV-bright regions ($\langle \log {\rm sSFR}\rangle = -9.86$) exhibit the opposite behavior ($\langle \Delta \log L_{\rm PAH}\rangle = -0.024$). We conclude that evolved stars contribute to PAH emission, and that such an effect, while relatively small, is systematic and should be considered whenever possible when using PAH emission to estimate SFR.

Having established that stellar mass contributes as a second parameter to the systematic scatter in the $L_{\rm PAH}-L_{\rm H\alpha}$ relation, we formally introduce $M_\star$ into a multiple regression analysis using the {\tt Python} package {\tt statsmodels} \citep{Seabold & Perktold 2010}:

\noindent\small\begin{align}\label{equ:PAH_Ha_Ms}
\begin{aligned}
&{\rm log}\,L_{\rm PAH} = (0.71\pm0.035)({\rm log}\, L_{\rm H\alpha} - 39.5)\ + \\&(0.28\pm0.032)({\rm log}\, M_{\star} - 6.5) + (40.47\pm0.056).
\end{aligned}
\end{align}

\normalsize\noindent
Compared with Equation~\ref{equ:SubRs}, Equation~\ref{equ:PAH_Ha_Ms} yields a slightly reduced total ($\epsilon_t = 0.17$\,dex) and intrinsic ($\epsilon_i = 0.11$\,dex) scatter. 

\subsection{Revised PAH-based SFR Calibration}\label{sec:sec3.2}

Figure~\ref{fig:Rev_SFR}a gives the conventional relationship between SFR and PAH luminosity for our sample. The best-fit linear regression (solid line), with $\epsilon_t = 0.21$\,dex and $\epsilon_i = 0.16$\,dex, is

\noindent\small\begin{align}\label{equ:SFR_PAH}
\begin{aligned}
&{\rm log}\,{\rm SFR} = (0.97\pm0.023)({\rm log}\, L_{\rm PAH} - 39.5) - (3.14\pm0.041).
\end{aligned}
\end{align}

\normalsize\noindent
The influence of evolved stars again becomes quite clear once we color code the points by their sSFR. The effect is further accentuated by superposing (in dashed line) the calibration of \cite{Xie & Ho 2019}\footnote{\cite{Xie & Ho 2019} define their integrated PAH emission over the region $5 - 15\ \mum$; their values need to be multiplied by a factor of 1.15 to be consistent with our definition, which is defined over $5 - 20\ \mum$. The SFRs of \cite{Xie & Ho 2019} also need to be multiplied by a factor of 0.68 due to be consistent with our choice of initial mass function.}, which has a higher zero point because their sample of massive SFGs has on average a factor of $\sim 2$ higher sSFR than our sample ($\langle \log {\rm sSFR}\rangle = -9.60$ versus $-9.93$).

Whenever possible, stellar mass should be included explicitly as an additional parameter to estimate SFRs, in order to remove the contribution of evolved stars to the excitation of PAH emission. We propose the following revised SFR calibration ($\epsilon_t = 0.21$\,dex, $\epsilon_i = 0.14$\,dex):

\noindent\small\begin{align}\label{equ:Ha_PAH_Ms}
\begin{aligned}
&{\rm log}\, {\rm SFR} = (1.08\pm0.052)({\rm log}\, L_{\rm PAH} - 39.5)\ - \\&(0.12\pm0.048)({\rm log}\, M_{\star} - 6.5) - (3.11\pm0.040).
\end{aligned}
\end{align}

\normalsize\noindent
Although the revised SFR calibration shows no signficant improvement in intrinsic scatter compared to the original calibration (Equation~\ref{equ:SFR_PAH}), the revised calibration provides more physically accurate and more robust SFR estimates, as evidenced by the systematic trend in Figure~\ref{fig:Rev_SFR}b. At a given observed PAH luminosity, the revised SFR estimates deviate systematically from the original values according to the sSFR (a proxy for the relative contribution from evolved stars). The revised calibration provides higher SFRs for regions with higher sSFR (blueish points), while the opposite situation holds for regions with lower sSFR (reddish points). This trend naturally arises if evolved stars contribution to PAH emission. The deviation on average is only at the level $\pm 0.05$~dex, but in the extreme it can reach $\pm 0.2$~dex.

\section{Discussion}\label{section:sec4}

\begin{figure}[t]
\center{\includegraphics[width=1\linewidth]{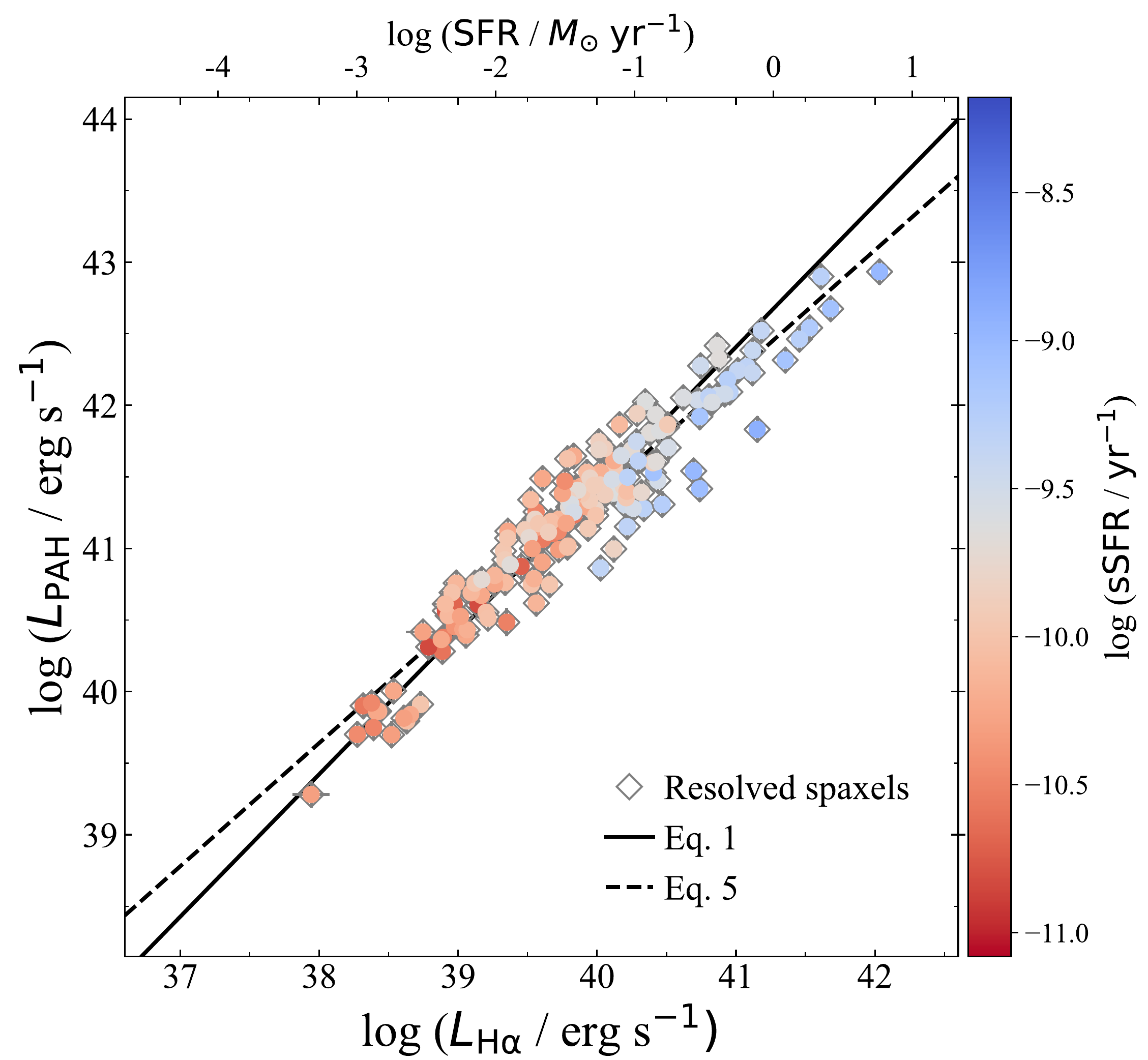}}
\caption{{The correlation between integrated PAH luminosity and extinction-corrected H$\alpha$ luminosity for the resolved, sub-kpc-scale spaxels from the central $\sim$30\arcsec$\times$50\arcsec\ regions of SINGS SFGs \citep{Zhang et al. 2022}. The black solid line is the best linear regression fit for the UV-bright (UVB) and UV-faint (UVF) regions (Equation~1), while the black dashed line is the best linear regression fit for the resolved spaxels (Equation~5).}\label{fig:PAH_SFR_CS}}
\end{figure}

\begin{figure*}[t]
\center{\includegraphics[width=0.95\linewidth]{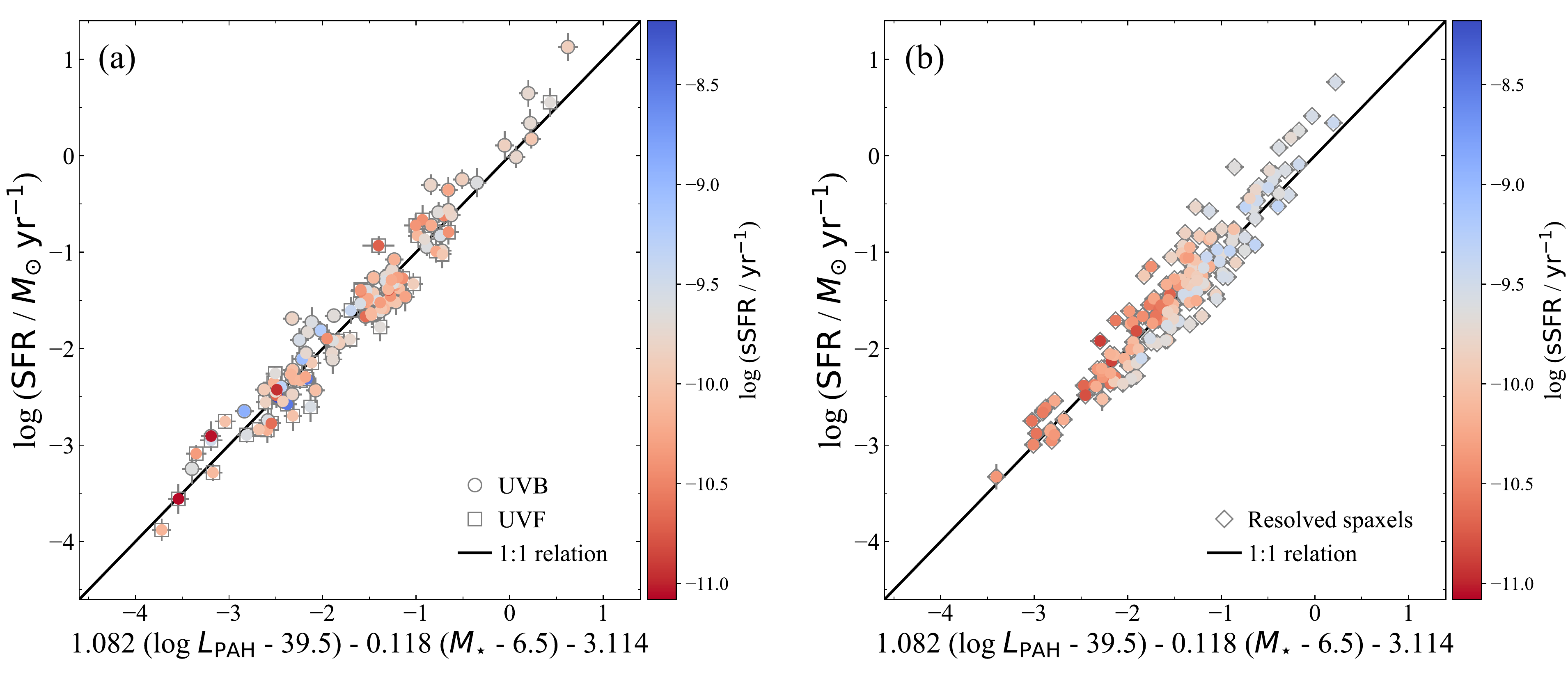}}
\caption{{The correlation between observed SFRs and revised SFRs (Equation~4) for (a) UV-bright (UVB) and UV-faint (UVF) regions and (b) resolved spaxels from the central $\sim$30\arcsec$\times$50\arcsec\ regions of SINGS SFGs \citep{Zhang et al. 2022}, color-coded according to the sSFR.}\label{fig:SFR_rSFR}}
\end{figure*}

A number of studies have noted that the correlation between PAH strength and SFR is sub-linear. For instance, \cite{Calzetti et al. 2005, Calzetti et al. 2007} and \cite{Wu et al. 2005} reported a best-fit slope of $\sim 0.9$ between the 8\,$\mum$ band PAH emission and different SFR indicators for resolved H\,{\small II} regions and SFGs. In their spatially resolved analysis of M\,51, \cite{Zhang et al. 2021} also found a sub-linear slope of 0.85 for the correlation between PAH emission and star formation indicators. A similar result appeared in \cite{Zhang et al. 2022}, who performed spatially resolved analysis of the SINGS sample, which includes the galaxies included in this study.

The sub-linear PAH--SFR correlation arises naturally if, as we contend, evolved stars contribute to PAH emission. We demonstrate this in Figure~\ref{fig:PAH_SFR_CS}, which repeats the $L_{\rm PAH}-L_{\rm H\alpha}$ relation of Figure~\ref{fig:PAH_Ha_SF_I_II}a, but now emphasizes instead the spaxels resolved on sub-kpc scales from the central $\sim$30\arcsec$\times$50\arcsec\ regions of the SINGS SFGs \citep{Zhang et al. 2022}. The corresponding linear regression fit ($\epsilon_t = 0.21$\,dex, $\epsilon_i = 0.19$\,dex),

\noindent\small\begin{align}\label{equ:ResSp}
\begin{aligned}
&{\rm log}\,L_{\rm PAH} = (0.86\pm0.022)({\rm log}\, L_{\rm H\alpha} - 39.5) + (40.93\pm0.018),
\end{aligned}
\end{align}

\normalsize\noindent
is clearly flatter (slope 0.86; dashed line) than that derived in Equation~1 for the much larger, spatially integrated UV-bright and UV-faint regions, whose slope is effectively unity (0.99, dashed line). Evidently, the chief culprit for the flatter correlation is the difference in relative contribution of evolved stars to PAH emission at the two ends of the SFR distribution. Unlike the UV-bright and UV-faint regions, which exhibit sSFR variation {\it perpendicular}\ to the correlation between PAH emission and SFR (Figure~\ref{fig:PAH_Ha_SF_I_II}a), the resolved spaxels show sSFR varying {\it along}\ the correlation. The increase of sSFR along the correlation elevates PAH emission at the low-SFR end, all the while diminishing PAH emission at the high-SFR end, the combination of which produces a flattening of the PAH--SFR correlation. While the destruction of PAHs within the resolved spaxels of the strongest SFR potentially can contribute to the flattening, this effect should play a secondary role here. In contrast to the sensitivity of the slope of the $L_{\rm PAH}-L_{\rm H\alpha}$ relation to the degree to which the galaxy is resolved, the differences disappear when we use the revised SFR estimates, which take into consideration the contribution of evolved stars (Figure~\ref{fig:SFR_rSFR}).

Given that the sum of the two coefficients in Equation~\ref{equ:PAH_Ha_Ms} is nearly unity, galaxies on the star-forming main sequence, whose SFR scales approximately linearly with stellar mass (e.g., \citealt{Brinchmann et al. 2004, Salim et al. 2007, Peng et al. 2010}), naturally present a linear correlation between PAH emission and SFR, if they are not biased by sample selection. Namely, the linear correlation between PAH and H$\alpha$ emission for UV-bright and UV-faint regions merely reflects the linear correlation between SFR and stellar mass. Similarly, the sub-linear correlation between PAH emission and SFR indicator implies a super-linear correlation between SFR and stellar mass (i.e., a sub-linear correlation between stellar mass and SFR). As the slope of the resolved star-forming main sequence spans a large range ($\sim 0.5-1.5$; \citealt{Maragkoudakis et al. 2017, Hall et al. 2018}), we argue that the correlation between SFR and PAH luminosity can be either linear or non-linear, depending on the sSFR, or the relative contribution of evolved stars, of the sample used for the empirical calibration. 

We close with a brief remark concerning the possible relevance of our results to the PAH deficit associated with metal-poor galactic environments (e.g., \citealt{Engelbracht et al. 2005, Jackson et al. 2006, Madden et al. 2006, Smith et al. 2007b, Wu et al. 2007, Khramtsova et al. 2013, Xie & Ho 2019}). The destruction of PAH molecules by the hard radiation field under low-metallicity conditions is widely invoked as the main reason for the observed deficit \citep{Jackson et al. 2006, Madden et al. 2006, Hunt et al. 2010, Khramtsova et al. 2013}. However, low-metallicity environments might also lead to the delayed formation of PAH molecules, if PAHs are produced by asymptotic giant branch stars \citep{Latter 1991, Dwek 1998}, which may not have accumulated in sufficient numbers in the young stellar population of low-metallicity environments \citep{Engelbracht et al. 2005, Wu et al. 2007, Galliano et al. 2008a}. Regardless of the mechanism responsible for the reduction of PAH content, the results of this work suggest that the characteristically lower fraction of evolved stars in low-metallicity environments should contribute, at least in part, to their observed PAH deficit.

\section{Summary}\label{section:sec5}

PAH emission, a commonly used indicator of star formation in galaxies, may be excited by both young and old stars. To assess the contribution of evolved stars to PAH emission, we perform a spatially resolved analysis of the nuclear and extra-nuclear regions of 33 nearby galaxies selected from the SINGS project. Employing the technique of \cite{Zhang et al. 2021, Zhang et al. 2022}, we successfully combine mapping-mode low-resolution IRS spectra with complementary near-IR and mid-IR images to derive robust meaurements of PAH luminosity and stellar mass. Together with SFRs obtained from narrowband H$\alpha$ images, we reinvestigate the correlation between PAH luminosity and SFR and its possible dependence on a second parameter. Classifying the level of star formation  activity with the intensity of UV emission measured from GALEX images, we show that UV-faint regions, which have characteristically lower sSFR and hence more prominent contribution from evolved stars, tend to exhibit stronger PAH emission at a given SFR than UV-bright regions.  We propose a revised calibration for estimating SFR based on both PAH luminosity and stellar mass. This new calibration, which explicitly takes into account the contribution of evolved stars to PAH emission, provides more accurate SFRs, especially for systems of low sSFR. The contribution of evolved stars to PAH emission naturally explains the well-documented sub-linear slope of the correlation between PAH emission and SFR, and it partly alleviates the PAH deficit observed in dwarf galaxies and other low-metallicity environments.

\acknowledgments
We thank the anonymous referee for constructive suggestions. This work was supported by the National Science Foundation of China (11721303, 11991052, 12011540375) and the China Manned Space Project (CMS-CSST-2021-A04, CMS-CSST-2021-A06).

\end{document}